\documentclass[prb,twocolumn]{revtex4}
\usepackage{color}
\usepackage{graphicx}

\begin{document}

\title{Spectroscopy of the transition-rate matrix for molecular junctions:\\
dynamics in the Franck-Condon regime}

\author{Agnieszka Donabidowicz-Kolkowska}
\affiliation{Institut f\"ur Theoretische Physik, Technische Universit\"at
Dresden, 01062 Dresden, Germany}

\author{Carsten Timm}
\email{carsten.timm@tu-dresden.de}
\affiliation{Institut f\"ur Theoretische Physik, Technische Universit\"at
Dresden, 01062 Dresden, Germany}

\date{August 10, 2012}

\begin{abstract}
The quantum master equation applied to electronic transport through nanoscopic
devices provides information not only on the stationary state but also on the
dynamics. The dynamics is characterized by the eigenvalues of the
transition-rate matrix, or generator, of the master equation. We propose to use
the spectrum of these eigenvalues as a tool for the study of nanoscopic
transport. We illustrate this idea by analyzing a molecular quantum dot with an
electronic orbital coupled to a vibrational mode, which shows the Franck-Condon
blockade if the coupling is strong. Our approach provides complementary
information compared to the study of observables in the stationary state.
\end{abstract}


\maketitle

\section{Introduction}

Recent progress in nanotechnology allows to fabricate transistors with a single
molecule forming the active region.\cite{Rat02,XuR06,ABV06,GRN07} The transport
properties of such devices have been studied extensively both experimentally and
theoretically. Theoretical approaches have been reviewed by Andergassen
\textit{et al.}\cite{Andergassen} and by Zimbovskaya and Pederson.\cite{ZiP11}

The transport properties of single-molecule devices are often dominated by
strong interactions. Electron-electron interaction, which leads to the Coulomb
blockade,\cite{Park,OOS07} and the coupling of electrons to vibrational
modes\cite{Park,Smit,Yu,OOS07} have to be taken into account.
We assume the nuclear motion to be slow on the time scale of
electronic transitions, which is the case for most, but not for
all, single-molecule devices studied so far.\cite{RLM10}
The electron-vibron coupling is then due to the
change of the equilibrium nuclear configuration with
the electronic state, i.e., the Franck-Condon effect:
Even if a molecule
is initially in a vibrational eigenstate, after an electronic transition it will
be in a superposition of vibrational eigenstates belonging
to the new electronic configuration. The probability to end up in any one of
them is determined by Franck-Condon matrix elements between vibrational
eigenstates for the old and new electronic
configurations.\cite{Fleb,Braig,KoO05,KRO05,KOA06} These matrix
elements can be very small if the equilibrium value of the relevant normal
coordinate changes strongly with the electronic state.
Consequently, the rates for electrons tunneling into or out of the molecule and
thus the current can be strongly suppressed---this is the Franck-Condon
blockade.\cite{KoO05,KRO05,KOA06,DGR06,Wegewijs,HuB09,Leturcq,CML10} The
dynamics is also
unusual in this regime: Electrons tunnel in avalanches separated by quiet time
intervals.\cite{KoO05,KRO05} The avalanche-type transport is self-similar on
intermediate timescales and also leads to a strong enhancement of the
zero-frequency Fano factor.\cite{KoO05,KRO05,KOA06}

Strong interactions and states far from equilibrium make the description of
molecular devices difficult in general.\cite{Andergassen,ZiP11}
We here focus on the case of weak hybridization between the molecule and the
leads. In this case, master-equation (ME)
approaches\cite{Braig,Mitra,KoO05,KOA06,DGR06,Wegewijs,Elste,HBT09,HuB09}
or the equivalent real-time diagrammatic approach\cite{ScS94,KSS95,Timm}
can be employed. In principle, the ME takes into account all interactions
in the molecule but requires an approximate treatment of the coupling to the
leads.

The simplest non-trivial version of the ME treats the hybridization to second
order (sequential tunneling) and neglects off-diagonal
components of the reduced density matrix in the eigenbasis of the
molecular Hamiltonian.\cite{Braig,Mitra,KoO05} The neglect of the off-diagonal
components is generally not justified if some of the eigenstates are
degenerate, since then the choice of eigenstates in the
degenerate subspaces is arbitrary. In the absence of spin-dependent terms in the
Hamiltonian, spin degeneracy is always present. It is then appropriate to
retain all \emph{secular} components of the reduced density
operator,\cite{Blum,Wegewijs,Koller} i.e., all diagonal matrix elements and
off-diagonal matrix elements between degenerate states.

Going beyond sequential tunneling, at fourth order one obtains cotunneling and
pair tunneling as well as a contribution to the life-time broadening of the
molecular levels.\cite{Koller} Koch \textit{et al.}\cite{KOA06} include
cotunneling but consider only the diagonal components of the reduced density
matrix. However, integrating out the off-diagonal terms of second order
generates contributions of fourth order to the transition rates between the
diagonal components, which are taken into account by Leijnse and
Wegewijs\cite{Wegewijs} as well as Koller \textit{et al.}\cite{Koller}

The dynamics of charge transport through molecules, in particular in the
Franck-Condon regime, have so far mostly been studied by considering the
current-noise
spectrum and the full counting statistics.\cite{KoO05,KRO05,KOA06,SKB12} For the
Franck-Condon regime, these studies have been augmented by real-time Monte
Carlo simulations.\cite{KoO05,KRO05,KOA06} Recently, Donarini \textit{et
al.}\cite{DYG12} have considered a harmonically varying bias voltage. Assuming
spinless electrons and employing the Markov and sequential-tunneling
approximations, they find hysteretic behavior of current and
charge.\cite{DYG12} In the present paper, we propose and
illustrate a different approach to the dynamics: We study the eigenvalue
spectrum of the transition-rate matrix appearing in the ME. Since
these eigenvalues describe the decay rates and oscillation frequencies of
\emph{deviations} of the reduced density matrix from the stationary solution,
they provide a complementary view of the dynamics. More specifically, we will
consider the part of the spectrum that is small in absolute value,
corresponding to states that decay or oscillate slowly. We use the well-studied
molecular transistor with a single vibrational mode as a testbed for our idea.

The remainder of this paper is organized as follows: In Sec.\ \ref{sec:theory}
the model is briefly introduced and the ME and the relevant
approximations are discussed. The tran\-si\-tion-rate matrix is
introduced and the physical interpretation of its eigenvalues is given. Results
for the eigenvalue spectra in various regimes are presented and discussed in
Sec.\ \ref{sec:results}. Section \ref{sec:conc} summarizes the main results and
draws some conclusions.

\section{Model and Method}
\label{sec:theory}

\subsection{Anderson-Holstein Hamiltonian}
\label{sus:model}

In the following we consider a device containing a single molecule coupled to
two electrodes. The device is described by the Anderson-Holstein Hamiltonian
$H=H_\mathrm{leads}+H_\mathrm{mol}+H_t$,\cite{GlS88,WJW88,Mitra,KoO05} where
\begin{eqnarray}
H_\mathrm{leads}=
  \sum_{\nu{\bf k}\sigma} \epsilon_{\bf k}\,
  c^\dagger_{\nu{\bf k}\sigma} c_{\nu{\bf k}\sigma}
\end{eqnarray}
refers to the noninteracting leads. For simplicitly, we assume that both
electrodes have identical band structures and a constant density of states. The
operator $c^\dagger_{\nu{\bf k}\sigma}$ creates an electron in lead $\nu=L,R$
with momentum $\mathbf{k}$ and spin $\sigma$. The molecule is described by
\begin{eqnarray}
H_\mathrm{mol} & = & \sum_{\sigma} \epsilon_d\,
  d^\dagger_{\sigma}d_{\sigma} + \frac{U}{2}\, \hat n_d(\hat n_d-1) \nonumber \\
& & {} + \hbar \omega_v\, \bigg(b^\dagger b + \frac{1}{2}\bigg)
  + \lambda \hbar\omega_v\,(b+b^\dagger)\,\hat n_d ,
\label{Hmol}
\end{eqnarray}
where $d^\dagger_\sigma$ creates an electron with spin $\sigma$ and energy
$\epsilon_d$ in the molecular orbital,
$\hat n_d = d_\uparrow^\dagger d_\uparrow + d_\downarrow^\dagger d_\downarrow$
is the corresponding number operator, and
$b^\dagger$ is the raising operator of a harmonic vibration mode.
In a break-junction device, the on-site energy can be tuned by
a gate voltage, which is absorbed into $\epsilon_d$. Finally, the tunneling
between the molecule and the leads is described by
\begin{eqnarray}
H_t = - \frac{1}{\sqrt{N}} \sum_{\nu{\bf k}\sigma} \left(
  t_\nu \,d_{\sigma}^\dagger c_{\nu{\bf k}\sigma}
  + \mathrm{H.c.} \right) ,
\label{Hhyb.1}
\end{eqnarray}
where $N\gg 1$ is the number of sites in one lead, which is often
absorbed into $t_\nu$.\cite{And61} We here assume that the
tunneling matrix elements $t_\nu$ are independent of momentum and spin in the
relevant energy range and that the contacts are symmetric.

\subsection{Master equation}
\label{sus:master}

The reduced density operator of the molecule is
\begin{eqnarray}
\rho_\mathrm{mol}(t) \equiv \text{Tr}_\mathrm{leads}\, \rho(t) ,
\label{u}
\end{eqnarray}
where $\rho(t)$ is the full density operator of the molecule and the leads and
the trace is over all many-particle states of the leads. The full density
operator satisfies the von Neumann equation
\begin{eqnarray}
\frac{d\rho}{dt}=-\frac{i}{\hbar}\, [H,\rho] .
\label{vonNeumann}
\end{eqnarray}
There are various ways to obtain a ME starting from Eq.\
(\ref{vonNeumann}).\cite{BrP02} Under the condition that the
system was in a product states with the leads in (separate) thermal equilibrium
at some initial time $t_0$, $\rho(t_0) = \rho_\mathrm{mol}(t_0) \otimes
\rho_\mathrm{leads}^0$, one can derive a ME that is local in
time.\cite{Nak58,Zwa60,ToM76,STH77,BrP02,Timm,Tim11}
This so-called time-convolutionless ME reads
\begin{eqnarray}
\frac{d\rho_\mathrm{mol}}{dt} = -\frac{i}{\hbar}\,
  [H_\mathrm{mol},\rho_\mathrm{mol}]
  - \mathcal{L}(t,t_0)\, \rho_\mathrm{mol} \equiv \mathcal{A}\,
  \rho_\mathrm{mol}
\label{TCL.1}
\end{eqnarray}
where the commutator induces the bare time evolution of the decoupled
molecule and $\mathcal{L}(t,t_0)$ is a linear superoperator describing the
coupling to the leads. The right-hand side of the ME is linear in
$\rho_\mathrm{mol}$ so that we can rewrite it as a product involving a
superoperator $\mathcal{A}$.

We write $\rho_\mathrm{mol}$ in the basis of eigenstates $|n,q \rangle$ to the
eigenvalues $E_{nq}$ of $H_\mathrm{mol}$, where $n\in
\{0,\uparrow,\downarrow,\uparrow\downarrow\}$ specifies the electronic state and
$q$ is the quantum number of the vibration. The eigenenergies are
\begin{eqnarray}
E_{nq} & = & \epsilon_d\, n_d + \frac{U}{2}\, n_d(n_d-1) \nonumber \\
&& {}+ \hbar\omega_v \bigg(q+\frac{1}{2}\bigg)
  - \lambda^2\hbar\omega_v\, n_d^2 \nonumber \\
& = & (\epsilon_d - \lambda^2\hbar\omega_v)\, n_d
  + \frac{1}{2}\, (U - 2\lambda^2\hbar\omega_v) \,
   n_d(n_d-1) \nonumber \\
&& {}+ \hbar\omega_v \bigg(q+\frac{1}{2}\bigg) ,
\label{eigen}
\end{eqnarray}
where $n_d(n)=0,1,2$ is the number of electrons in the electronic state $n$.
The matrix elements of the reduced density operator in this basis are denoted by
\begin{eqnarray}
\rho^{nn'}_{qq'} \equiv \langle n,q|\, \rho_\mathrm{mol}\, |n',q'\rangle .
\end{eqnarray}
The ME (\ref{TCL.1}) written in components reads
\begin{eqnarray}
\frac{d}{dt}\, \rho^{nn'}_{qq'} & = & -\frac{i}{\hbar}\,
  \big(E_{nq}-E_{n' q'} \big)\, \rho^{n n'}_{q q'} \nonumber \\
&& {}- \sum_{n''q''n'''q'''}
  \mathcal{L}^{nn',n''n'''}_{qq',q''q'''} \rho^{n''n'''}_{q''q'''} .
\label{TCL.2}
\end{eqnarray}
Expressed precisely, our goal is to find the eigenvalue
spectrum of the superoperator $\mathcal{A}$ in the ME (\ref{TCL.1}). The
physical interpretation of the eigenvalues becomes clear if one inserts the
ansatz
\begin{eqnarray}
\rho_\mathrm{mol}(t) = e^{\alpha t}\, \zeta_\alpha
\label{rhoansatz.2}
\end{eqnarray}
into Eq.\ (\ref{TCL.1}). Here, $\alpha$ is a complex number and $\zeta_\alpha$
is an operator on the molecular Fock space. This leads to the eigenvalue
equation
\begin{eqnarray}
\mathcal{A}\, \zeta_\alpha = \alpha\, \zeta_\alpha .
\label{evaleq.1}
\end{eqnarray}
The ansatz (\ref{rhoansatz.2}) indeed solves the ME if
$\alpha$ is an eigenvalue of $\mathcal{A}$. Then,
$\mathrm{Re}\, \alpha$ is the negative of the decay rate of the corresponding
solution, while $\mathrm{Im}\, \alpha$ is the angular frequency of its
oscillations. Evidently, a vanishing eigenvalue $\alpha=0$ corresponds to a
stationary state. The stationary density operator is thus the right eigenvector
$\zeta_0$ if we impose the normalization condition $\mathrm{Tr}\,\zeta_0=1$.

Since the ME (\ref{TCL.2}) has to preserve the trace of
$\rho_\mathrm{mol}$, it must satisfy
\begin{eqnarray}
0 = \sum_{nq} \frac{d\rho^{nn}_{qq}}{dt}
  = - \sum_{nq} \sum_{n''q''n'''q'''}
  \mathcal{L}^{nn,n''n'''}_{qq,q''q'''} \rho^{n''n'''}_{q''q'''}
\end{eqnarray}
for all $\rho_\mathrm{mol}$. Therefore, $\eta_0$ with components
$\eta^{nn'}_{0,qq'} \equiv \delta_{nn'} \delta_{qq'}$ is a \emph{left}
eigenvector of
$\mathcal{A}$ to the eigenvalue zero. This proves that at least one stationary
state exists. This solution is unique if the
system is ergodic in the sense that every state can be reached from every other
state by a finite number of transitions.\cite{Bre99,Muk00,ZiS06,timm2} This
condition is satisfied by our model for non-zero temperature. Thus there is
exactly one eigenvalue $\alpha=0$.

What is the meaning of the other right eigenvectors $\zeta_\alpha$ for
$\alpha\neq 0$? These eigenvectors are not well-formed density matrices since
they have zero trace. This follows from the fact that $\eta_0$ is a left
eigenvector to eigenvalue zero. Since the left and right eigenvectors to
different eigenvalues are orthogonal, one has for all right
eigenvectors $\zeta_\alpha$ to non-vanishing eigenvalues
\begin{eqnarray}
0 = \mathrm{Tr}\, \eta_0^\dagger \zeta_\alpha
  = \sum_{nqn'q'} \big(\eta^{n'n}_{0,q'q}\big)^*\,
    \zeta^{nn'}_{\alpha,qq'}
  = \sum_{nq} \zeta^{nn}_{\alpha,qq} .
\end{eqnarray}
Thus one cannot interpret $\zeta_\alpha$ as a density matrix. However, linear
combinations of the form
\begin{eqnarray}
\rho_\mathrm{mol}(t) = \zeta_0 + \sum_{\alpha\neq 0} c_\alpha e^{\alpha t}
  \zeta_\alpha ,
\end{eqnarray}
with constants $c_\alpha$ chosen such that $\rho_\mathrm{mol}$ is hermitian,
have unit trace and statisfy the ME (\ref{TCL.1}). As long as
$\rho_\mathrm{mol}$ is a positive matrix, it is a permissable time-dependent
density matrix. Hence, the eigenvectors
$\zeta_\alpha$ for $\alpha\neq 0$ describe \emph{deviations} of possible
density matrices from the stationary state. Their time dependence is governed by
the eigenvalue $\alpha$. One can show that all eigenvalues $\alpha\neq 0$ have
negative real parts,\cite{BeP79} i.e, they describe deviations from
the stationary state that decay for $t\to\infty$.

In practice, approximations are needed to obtain the superoperator
$\mathcal{A}$. We employ the
sequential-tunneling and secular approximations. While these are
standard for the study of stationary states, we have to show that they
are justified for our purpose of obtaining the eigenvalue spectrum.
We assume weak coupling between molecule and leads and
expand the right-hand side of Eq.\ (\ref{TCL.1}) in powers of $t_{L,R}$. For the
tunnel Hamiltonian $H_t$, only even powers of $t_{L,R}$ enter. One can thus
write
\begin{eqnarray}
\frac{d\rho_\mathrm{mol}}{dt} = \sum_{n=0}^\infty
  \mathcal{A}^{(2n)}\, \rho_\mathrm{mol} .
\label{TCL.3}
\end{eqnarray}
Here, $\mathcal{A}^{(0)}
\rho_\mathrm{mol} = -(i/\hbar)\,[H_\mathrm{mol},\rho_\mathrm{mol}]$ is the bare
time evolution
from Eqs.\ (\ref{TCL.1}) and (\ref{TCL.2}). Equation (\ref{TCL.2}) shows that
$\mathcal{A}^{(0)}$ is diagonal in the basis $\{|n,q\rangle\langle n',q'|\}$.
We split the reduced density
operator into a secular part $\rho_s$ and a non-secular part $\rho_n$ and
expand both,
\begin{eqnarray}
\rho_{s,n} = \sum_{n=0}^\infty \rho^{(2n)}_{s,n} .
\end{eqnarray}
This expansion allows us to write down the ME order by order in $t_{L,R}^2$.

First, we discuss the stationary state, for which the left-hand side
of the ME (\ref{TCL.1}) vanishes. At order zero, we obtain
\begin{eqnarray}
\mathcal{A}^{(0)} \rho^{(0)}_{s}
  + \mathcal{A}^{(0)} \rho^{(0)}_{n} = 0 .
\end{eqnarray}
Since $\mathcal{A}^{(0)}$ is diagonal and according to Eq.\ (\ref{TCL.2}) gives
zero when acting on the secular part $\rho^{(0)}_{s}$, we obtain
$\mathcal{A}^{(0)} \rho^{(0)}_{n} = 0$. This implies that $\rho^{(0)}_{n} = 0$,
since $\mathcal{A}^{(0)}$
multiplies any non-secular component of the reduced density operator by a
non-vanishing factor. At second order we thus find
\begin{eqnarray}
\lefteqn{ \mathcal{A}^{(2)} \rho^{(0)}_{s}
  + \mathcal{A}^{(2)} \rho^{(0)}_{n}
  + \mathcal{A}^{(0)} \rho^{(2)}_{s}
  + \mathcal{A}^{(0)} \rho^{(2)}_{n} } \nonumber \\
&& = \mathcal{A}^{(2)} \rho^{(0)}_{s}
  + \mathcal{A}^{(0)} \rho^{(2)}_{n} = 0 .\hspace{5em}
\label{MEO2}
\end{eqnarray}
The leading secular components are thus of order zero, while
the non-secular components are at least of order two.
From the ME up to second order in $t_{L,R}$, one cannot obtain the
second-order corrections to the secular part, $\rho^{(2)}_{s}$,
this requires to go to fourth order.\cite{Wegewijs,Koller} The
leading-order stationary density operator is thus the solution of
\begin{eqnarray}
\mathcal{A}^{(2)} \rho^{(0)}_{s} = 0
\end{eqnarray}
with the non-secular components vanishing, $\rho^{(0)}_{n} = 0$.

Next, we turn to the dynamics. In the ME (\ref{TCL.2}), the time
derivative of all non-secular components of $\rho_\mathrm{mol}$ picks up
imaginary factors $-(i/\hbar)\,(E_{nq}-E_{n'q'})$. They are large compared to
the typical scale introduced by the coupling since we assume the coupling
to be weak and exclude accidental near-degeneracies. To put it differently, the
zero-order superoperator $\mathcal{A}^{(0)}$ has eigenvalues
$\alpha^{(0),nn'}_{qq'} = -(i/\hbar)\,(E_{nq}-E_{n'q'})$ to eigenstates
$|n,q\rangle\langle n',q'|$.
These eigenvalues are zero for secular components and large and purely imaginary
for non-secular components. As long as the perturbative expansion in $t_{L,R}$
is justified, the full superoperator $\mathcal{A}$ will have
eigenvalues close to the zero-order ones, thus some will be small in absolute
value, while the rest will have a large imaginary part. We are interested in
the part of the spectrum with small absolute value. This is the part lacking
large imaginary parts from order zero.
Consequently, to find the small eigenvalues, one has to consider the secular
sector. To leading order, the eigenvalues are then given by the second-order
superoperator $\mathcal{A}^{(2)}$. Consequently, we have to solve the ME
\begin{eqnarray}
\frac{d\rho_s}{dt} = \mathcal{A}^{(2)} \rho_s
\label{TCL.4}
\end{eqnarray}
for the secular density operator $\rho_s$.
The derivation of $\mathcal{A}^{(2)}$ is standard and we
omit the details.\cite{Blum,Mitra,KoO05,ElT05,TiE06,KOA06,Timm} Assuming that at
some initial time $t_0$ the molecule and the leads are in a product state with
the leads in separate equilibrium, $\rho(t_0)= \rho_\mathrm{mol}(t_0)
\otimes \rho^0_\mathrm{leads}$, and taking the limit $t_0\to-\infty$, one
obtains to second order in $t_{L,R}$,
\begin{eqnarray}
\lefteqn{ \frac{d\rho_\mathrm{mol}}{dt} = - \frac{i}{\hbar}\,
  [H_\mathrm{mol},\rho_\mathrm{mol}(t)] } \nonumber \\
&& {}- \frac{1}{\hbar^2} \int_{0}^{\infty} d\tau\: \text{Tr}_\mathrm{leads}\,
  \big[H_t, [e^{-i(H_\mathrm{leads}+H_\mathrm{mol})\tau/\hbar} \nonumber
  \\
&& {}\times H_t\,
  e^{i(H_\mathrm{leads}+H_\mathrm{mol})\tau/\hbar}, \rho_\mathrm{mol}(t) \otimes
  \rho^0_\mathrm{leads}] \big] .\quad
\label{milo}
\end{eqnarray}
We expand the nested commutators and take the trace over the lead degrees of
freedom using
\begin{eqnarray}
\text{Tr}_\mathrm{leads}\, \rho^0_\mathrm{leads} c^\dagger_{\nu\mathbf{k}\sigma}
  c_{\nu'\mathbf{k}'\sigma'} & = & \delta_{\nu\nu'}
  \delta_{\mathbf{kk}'} \delta_{\sigma\sigma'} f(\xi_{\nu\mathbf{k}}) , \\
\text{Tr}_\mathrm{leads}\, \rho^0_\mathrm{leads} c_{\nu\mathbf{k}\sigma}
  c^\dagger_{\nu'\mathbf{k}'\sigma'} & = & \delta_{\nu\nu'}
  \delta_{\mathbf{kk}'} \delta_{\sigma\sigma'} [1-f(\xi_{\nu\mathbf{k}})]
  , \qquad
\end{eqnarray}
where $\xi_{\nu\mathbf{k}} \equiv \epsilon_{\mathbf{k}} - \mu_\nu$, $\mu_\nu$
is the chemical potential in lead $\nu$, and $f(\xi)$ is the Fermi function. The
chemical potentials in the left and right leads satisfy $\mu_R - \mu_L = eV$. We
assume that the device is symmetric, i.e., $\mu_L = -\mu_R = -eV/2$.
In the basis of eigenstates $|n,q\rangle$, the ME has the form
\begin{eqnarray}
\lefteqn{ \frac{d \rho^{n n'}_{q q'}}{dt} = -\frac{i}{\hbar}\,
  \big(E_{nq}-E_{n' q'} \big)\, \rho^{n n'}_{q q'}
  - \sum_{n''q''} R^{nn''}_{q q''}\rho^{n'' n'}_{q'' q'} } \nonumber \\
&& {}- \sum_{n''q''} \big( R^{n'n''}_{q'q''} \big)^* \rho^{n n''}_{q q''}
  + \!\! \sum_{n'' n''' q'' q'''} \! R^{nn'',n'''n'}_{q q'',q''' q'}
  \rho^{n'' n'''}_{q'' q'''} . \nonumber \\[-2ex]
&& {}
\label{masterequation}
\end{eqnarray}
Since we are only interested in the secular sector, we can drop the first term
on the right-hand side, which corresponds to $\mathcal{A}^{(0)}$.
The expressions for the rates $R$ also simplify in this
sector. The rates appropriate for secular $\rho_\mathrm{mol}$ read
\begin{widetext}
\begin{eqnarray}
R^{nn'}_{qq'} & = & \frac{\Gamma}{2} \sum_{n''q''}
  \bigg[ f\bigg(E_{n''q''}-E_{nq}-\frac{eV}{2}\bigg)
  + f\bigg(E_{n''q''}-E_{nq}+\frac{eV}{2}\bigg) \bigg] \nonumber \\
&& {}\times \bigg( \sum_\sigma D^\sigma_{nn''} D^{\dagger\sigma}_{n''n'}
  F_{qq''} F^\dagger_{q''q'}
  + \sum_\sigma D^{\dagger\sigma}_{nn''} D^\sigma_{n''n'}
  F^\dagger_{qq''} F_{q''q'} \bigg) ,
\label{R2} \\
R^{nn'',n'''n'}_{qq'',q'''q'} & = & \Gamma\,
  \bigg[ f\bigg(E_{nq}-E_{n''q''}-\frac{eV}{2}\bigg)
  + f\bigg(E_{nq}-E_{n''q''}+\frac{eV}{2}\bigg) \bigg] \nonumber \\
&& {}\times \bigg( \sum_\sigma D^\sigma_{nn''} D^{\dagger\sigma}_{n'''n'}
  F_{qq''} F^\dagger_{q'''q'}
  + \sum_\sigma D^{\dagger\sigma}_{nn''} D^\sigma_{n'''n'}
  F^\dagger_{qq''} F_{q'''q'} \bigg) ,
\label{R4}
\end{eqnarray}
\end{widetext}
where the rate $\Gamma\equiv 2\pi N_{L,R} |t_{L,R}|^{2}/\hbar$ describes the
coupling to the electrodes with densities of states $N_{L,R}$, which are assumed
to be constant, and
$D^{\sigma}_{nn'} \equiv \langle n|d_{\sigma}|n'\rangle$,
$D^{\dagger\sigma}_{nn'} \equiv \langle n|d^\dagger_\sigma|n'\rangle$
are matrix elements of the electronic operators. The
Franck-Condon matrix elements $F_{qq'}\equiv \langle
q|e^{-\lambda(b^\dagger-b)}|q'\rangle$
read explicitly\cite{Mitra,KoO05,KOA06,Wegewijs}
\begin{eqnarray}
F_{qq'} & = & \sqrt{\frac{q_<!}{q_>!}}\, \lambda^{q_>-q_<}\,
e^{-\lambda^2/2}\,
  L^{q_>-q_<}_{q_<}(\lambda^2) \nonumber \\
&& {}\times \left\{\begin{array}{ll}
    (-1)^{q-q'} & \mbox{for $q\ge q'$,} \\[0.7ex]
    1 & \mbox{for $q<q'$,}
  \end{array}\right.
\end{eqnarray}
and $F^\dagger_{qq'} \equiv (F_{q'q})^* = F_{q'q}$. Here, $q_<\equiv\min(q,q')$,
$q_>\equiv\max(q,q')$, and $L^i_j(x)$ are generalized Laguerre polynomials.

In Eqs.\ (\ref{R2}) and (\ref{R4}), the matrix elements
$D^{\sigma}_{nn'}$ and $D^{\dagger\sigma}_{nn'}$ always appear in combinations
corresponding to the creation and annihilation of electrons of the \emph{same}
spin $\sigma$. This leads to the
vanishing of the rates for certain combinations of electronic states.
Furthermore, the only off-diagonal secular components of $\rho_\mathrm{mol}$ are
$\rho^{\uparrow\downarrow}_{qq}$ and $\rho^{\downarrow\uparrow}_{qq}$ for all
$q$. Thus, the rates relevant for the secular sector simplify to
\begin{widetext}
\begin{eqnarray}
R^{nn'}_{qq} & = & \delta_{nn'}\, \frac{\Gamma}{2} \sum_{n''q''}
  \bigg[ f\bigg(E_{n''q''}-E_{nq}-\frac{eV}{2}\bigg)
  + f\bigg(E_{n''q''}-E_{nq}+\frac{eV}{2}\bigg) \bigg] \nonumber \\
&& {}\times \bigg( \sum_\sigma |D^\sigma_{nn''}|^2 |F_{qq''}|^2
  + \sum_\sigma |D^\sigma_{n''n}|^2 |F_{q''q}|^2 \bigg) ,
\label{R2a} \\
R^{nn'',n'''n'}_{qq'',q''q} & = & \delta_{nn'} \delta_{n''n'''}\, \Gamma\,
  \bigg[ f\bigg(E_{nq}-E_{n''q''}-\frac{eV}{2}\bigg)
  + f\bigg(E_{nq}-E_{n''q''}+\frac{eV}{2}\bigg) \bigg] \nonumber \\
&& {}\times \bigg( \sum_\sigma |D^\sigma_{nn''}|^2 |F_{qq''}|^2
  + \sum_\sigma |D^\sigma_{n''n}|^2 |F_{q''q}|^2 \bigg) .
\label{R4a}
\end{eqnarray}
\end{widetext}
Note that principal-value integrals, which plague the ME for the
full reduced density operator, cancel in the secular sector. Also, the
rates in Eqs.\ (\ref{R2a}) and (\ref{R4a}) are real. The diagonal components of
the ME (\ref{masterequation}) then simplify to
\begin{eqnarray}
\frac{d\rho^{nn}_{qq}}{dt} = -2\, R^{nn}_{qq} \, \rho^{nn}_{qq}
  + \sum_{n''q''} R^{nn'',n''n}_{qq'',q''q}\, \rho^{n''n''}_{q''q''} ,
\label{ME.5a}
\end{eqnarray}
while for the off-diagonal secular components we obtain
\begin{eqnarray}
\frac{d\rho^{nn'}_{qq}}{dt} = - (R^{nn}_{qq} + R^{n'n'}_{qq})\,
  \rho^{nn'}_{qq} .
\label{ME.5b}
\end{eqnarray}
The equations for the diagonal and the off-diagonal components thus decouple and
the off-diagonal components exhibit simple exponential decays.

Averages of local observables such as the electron number in the molecule can
be obtained directly from $\rho^{nn'}_{qq'}$. The current requires a
different approach. In the following, we consider a charge current flowing from
left to right as
positive. The operator $\hat I_\nu$ for the current between the lead $\nu=L,R$
and the molecule is
\begin{eqnarray}
\hat I_\nu = \nu e\, \frac{d}{dt}\, \hat N_\nu = \nu e\,\frac{i}{\hbar}\,
  [H,\hat N_\nu] = i\,\frac{\nu e}{\hbar}\, [H_t^\nu,\hat N_\nu] ,
\label{currentoperator}
\end{eqnarray}
where the numerical value of $\nu$ is $+1$ ($-1$) for the left (right) lead,
$-e$ is the charge of the electron, $\hat N_\nu$ is the
number operator of electrons in lead $\nu$, and $H_t^\nu$ is the part of the
tunneling Hamiltonian $H_t$ involving lead $\nu$. The current is then
$I_\nu = \text{Tr}\, \hat I_\nu\, \rho$,
where $\text{Tr}$ is the trace over the full Fock space. Under the same
assumptions as used for the rates above one obtains
\begin{widetext}
\begin{eqnarray}
I_\nu & = & \nu e \Gamma \sum_{nqn'q'} \sum_{n''q''}
  \bigg[ f\bigg(E_{n''q''}-E_{nq}-\nu\,\frac{eV}{2}\bigg)
    \sum_\sigma D^{\dagger\sigma}_{n'n''} D^\sigma_{n''n}
    F^\dagger_{q'q''} F_{q''q} \nonumber \\
&& {}- f\bigg(E_{n''q''}-E_{nq}+\nu\,\frac{eV}{2}\bigg)
    \sum_\sigma D^\sigma_{n'n''} D^{\dagger\sigma}_{n''n}
    F_{q'q''} F^\dagger_{q''q} \bigg]\, \rho^{nn'}_{qq'} ,
\end{eqnarray}
\end{widetext}
where the sum over $n$, $q$, $n'$, $q'$ only runs over secular components.

For the numerical calculations, we cut off the ladder of harmonic-oscillator
states so that $0\le q\le q_\mathrm{max}$. Then the dimension of the molecular
Fock space is $4(q_\mathrm{max}+1)$. The secular part of $\rho_\mathrm{mol}$
has $6(q_\mathrm{max}+1)$ components---the $4(q_\mathrm{max}+1)$
diagonal ones
and $2(q_\mathrm{max}+1)$ off-diagonal ones of the form
$\rho^{\uparrow\downarrow}_{qq}$ and
$\rho^{\downarrow\uparrow}_{qq}$. In the secular sector, the density operator
can thus be represented by a $6(q_\mathrm{max}+1)$-component vector and the
superoperator $\mathcal{A}^{(2)}$ by a real $6(q_\mathrm{max}+1) \times
6(q_\mathrm{max}+1)$ \emph{transition-rate matrix}.

It is a common problem of the ME approach that the matrix $\mathcal{A}^{(2)}$
can be ill-conditioned, i.e., the ratio of the largest to the smallest
eigenvalue can be very large. This is here due to the Fermi functions and
Franck-Condon matrix elements in Eqs.\
(\ref{R2a}) and (\ref{R4a}), which span many orders of magnitude. For this
reason, black-box diagonalization routines often fail to
distinguish the true stationary state from eigenvectors to very small
eigenvalues. To overcome this difficulty, we use Mathematica\cite{Mathematica}
to solve the eigenvalue problem with high precision. We adapt the number of
digits in the calculation such that it is larger than the $L^\infty$ condition
number of the matrix by at least $12$, which should give results for the
eigenvalues and eigenvectors with on the order of $12$ significant digits.

Before analyzing eigenvalue spectra in the next section, we comment on their
dependence on the cutoff $q_\mathrm{max}$. We find that adding highly excited
vibrational states only adds eigenvalues to the middle and upper part of the
spectrum. If $q_\mathrm{max}$ is not too small, the spectrum at small
magnitudes does not change significantly with $q_\mathrm{max}$. This is
plausible since adding highly excited vibrational states should not
introduce additional slow relaxation channels. We use $q_\mathrm{max}=30$,
unless noted otherwise.

It would be possible extend the analysis to higher
orders in the tunneling amplitudes $t_{L,R}$. This would require to solve the
eigenvalue equation (\ref{evaleq.1}) perturbatively. Compared to
time-independent perturbation theory for \emph{Hamiltonian} systems, this is
more complicated since the superoperator $\mathcal{A}$ is not
hermitian.\cite{StW72} In this work, we obtain the eigenvalues $\alpha$ up to
second order. The next non-vanishing contribution is of fourth
order. At this order, the restriction to the secular sector is not possible, we
require $\mathcal{A}^{(2)}$ for all secular and non-secular states, and we also
need $\mathcal{A}^{(4)}$ within the secular sector. The stationary solution at
this order has been studied before.\cite{Wegewijs,Koller} We leave the spectral
analysis for a future work.

\section{Results}
\label{sec:results}

In the present section, we present results for the eigenvalue spectrum in the
regimes of the transmitting quantum dot, the Coulomb blockade, and
the Franck-Condon blockade. It is shown that the spectrum differs qualitatively
between these cases and is characteristic for each. We compare the
information that can be gained more conventionally from observables such as
current and charge in the stationary state.

\begin{figure}[tbh]
\centerline{\includegraphics[width=3.60in,clip]{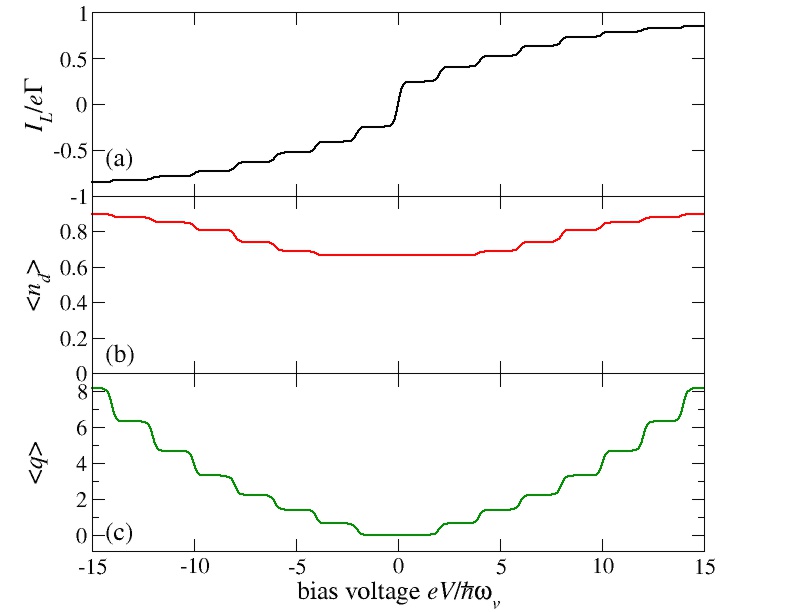}}
\caption{(Color online) (a) Current, (b) electronic occupation number, and (c)
excitation of the harmonic oscillator as functions of bias voltage $eV$ for
relatively small electron-vibron coupling $\lambda=1$, on-site energy
$\epsilon_d=1$, Hubbard interaction $U=6$, and thermal energy
$k_BT=0.05$. All energies are given in units of the vibron energy
$\hbar\omega_v=1$.}
\label{uli1}
\end{figure}

\begin{figure}[tbh]
\centerline{\includegraphics[width=3.40in,clip]{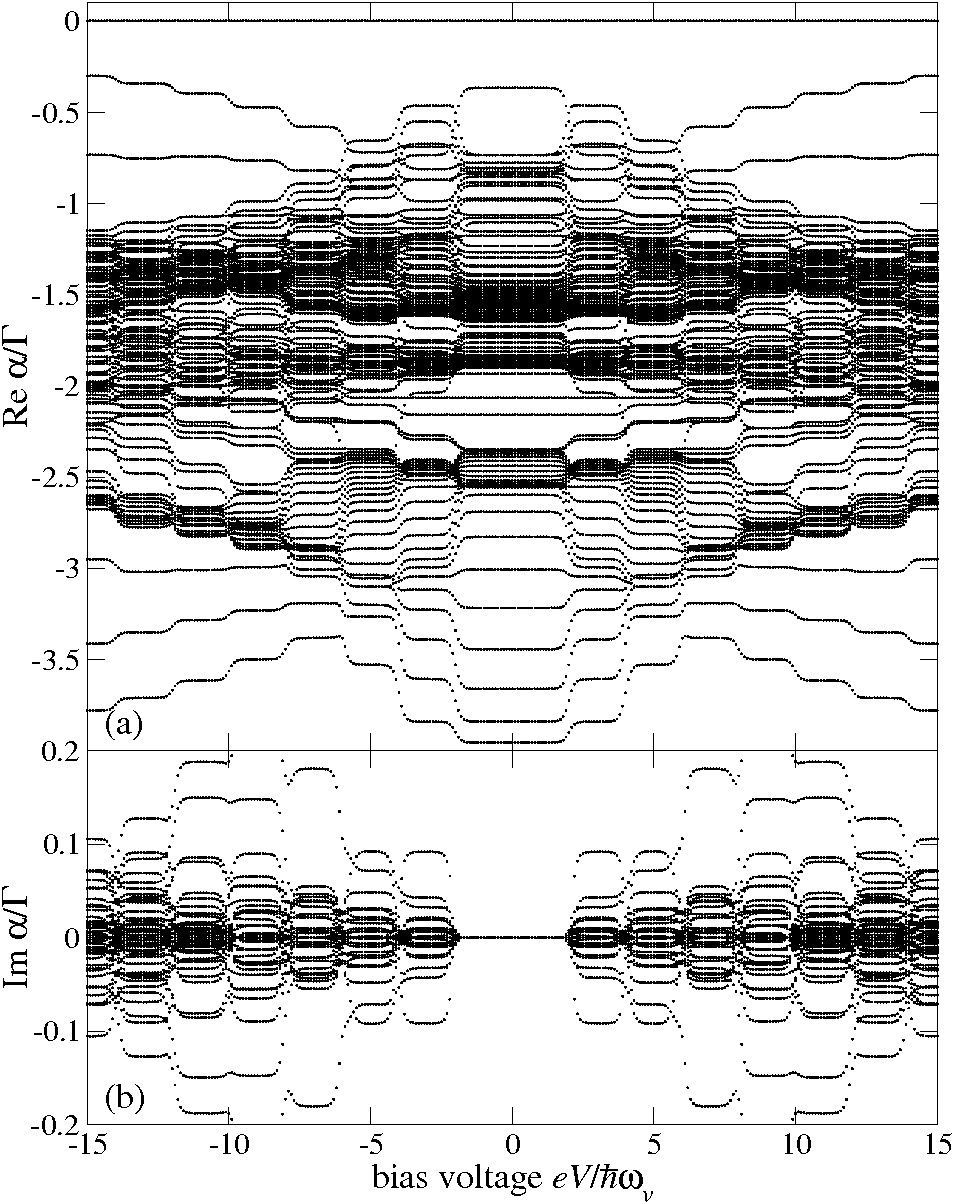}}
\caption{(a) Real and (b) imaginary parts of the eigenvalues of the
transition-rate matrix as functions of the applied bias voltage $eV$. The
parameters are the same as in Fig.\ \ref{uli1}.}
\label{fig1}
\end{figure}

\subsection{Transmitting regime}

First, we consider the transmitting molecular device. We tune the effective
on-site energy $\epsilon_d-\lambda^2\hbar\omega_v$ in Eq.\ (\ref{eigen}) to
zero so that resonant tunneling is possible at vanishing bias voltage. Figure
\ref{uli1}
shows the current $I_L$, the electronic occupation number $\langle n_d\rangle$,
and the vibron excitation $\langle q\rangle$ as functions
of the bias voltage in the stationary state for relatively small electron-vibron
coupling $\lambda=1$. The current increases with
characteristic steps.\cite{Braig,KoO05,Mitra} There is a step at zero bias
since the device is on resonance. The steps at non-zero bias result from
inelastic tunneling under excitation of $1,2,\ldots$ vibron
quanta,\cite{Braig,KoO05,Mitra} in agreement with the observed average
excitation $\langle q\rangle$. The average electronic occupation changes only
weakly and mostly above an energy scale on the order of $U$. This weak
dependence is due to a small admixture of doubly occupied states at higher bias
voltages.

The real and imaginary parts of the eigenvalues $\alpha$ of the transition-rate
matrix are shown in Fig.\ \ref{fig1} for the same parameters. The eigenvalue
zero is always present, as it must be. The real and imaginary parts of the other
eigenvalues reflect the step positions from Fig.\ \ref{uli1}, except for the
zero-bias step. Note that at any voltage, most eigenvalues have vanishing
imaginary parts. The eigenvalues with non-vanishing imaginary part form
complex-conjugate pairs,
since the transition-rate matrix is real. The non-vanishing
imaginary parts are typically small compared to the real parts. This means
that the decay time of the corresponding deviations from the stationary state
is much shorter than their oscillation period. Similar behavior is found for
randomly distributed transition rates, where it is essentially a consequence
of different scaling of the real and imaginary parts with the dimension of the
molecular Fock space.\cite{timm2}

At zero bias, all eigenvalues are real. This has to be the case since for
$V=0$ the molecule is coupled
to an equilibrium bath and all transition rates satisfy
detailed balance.\cite{Bre99,Muk00,ZiS06} This is easily confirmed by checking
that the non-vanishing rates in Eq.\ (\ref{ME.5a}) satisfy
$R^{n'n,nn'}_{q'q,qq'}/R^{nn',n'n}_{qq',q'q} = e^{(E_{nq}-E_{n'q'})/k_BT}$ at
$V=0$. Then for the diagonal components, the transition-rate matrix
$\mathcal{A}$ can be written in the form
\begin{eqnarray}
\mathcal{A}_{ij} = \left\{\begin{array}{ll}
  R^0_{ij}\, e^{\beta(E_j-E_i)/2} & \mbox{for $i\neq j$}, \\[1ex]
  -\sum_{k\neq i} R^0_{ki}\, e^{\beta(E_i-E_k)/2} & \mbox{for $i=j$,}
  \end{array}\right.
\end{eqnarray}
where $\beta=1/k_BT$ is the inverse temperature and $R^0_{ij}=R^0_{ji}$.
Introducing the diagonal superoperator $\mathcal{O}$ with components
$\mathcal{O}_{ij} = \delta_{ij}\, e^{\beta E_i/2}$, one easily sees that
$\mathcal{O}\mathcal{A}\mathcal{O}^{-1}$ is real and symmetric and therefore
has only real eigenvalues. Since this is a similarity transformation,
$\mathcal{A}$ has the same real eigenvalues.\cite{Bre99,Muk00,BYP05} The
previous argument only applies to the diagonal components. However, the
off-diagonal secular components show a simple exponential decay anyway, as
expressed by Eq.\ (\ref{ME.5b}).

Concerning our goal of characterizing different regimes in terms of their
eigenvalue spectra, the crucial observation is that the spectrum shows a
clear gap in the real part. Thus there are no slow modes---all deviations
from the stationary state decay with rates that are on the order of the
characteristic rate $\Gamma$.

It is interesting to analyze the character of the stationary state and of the
deviations that decay most slowly. At $V=0$, the stationary (equilibrium)
state is a mixture of the microstates $|n,q\rangle = |0,0\rangle,\,
|{\uparrow},0\rangle,\, |{\downarrow},0\rangle$ with equal probabilities, except
for exponentially small thermal occupations of higher-\textit{q}
and doubly occupied states, see Figs.\ \ref{uli1}(b) and \ref{uli1}(c).

As discussed, the eigenvectors to non-vanishing eigenvalues represent
deviations from the stationary state. The components $\zeta^{nn'}_{\alpha,qq}$
with the largest magnitudes characterize the microstates that have the largest
weight in a given deviation. At $V=0$, we find that the eigenvalue with the
smallest non-vanishing magnitude is actually threefold degenerate---it
corresponds to three linearly independent deviations.
The corresponding subspace is spanned by the
hermitian matrices $|{\uparrow},0\rangle\langle\uparrow,0|
-|{\downarrow},0\rangle\langle{\downarrow,0}|$,
$|{\uparrow},0\rangle\langle\downarrow,0|
+|{\downarrow},0\rangle\langle\uparrow,0|$ and
$-i\,|{\uparrow},0\rangle\langle\downarrow,0|
+i\,|{\downarrow},0\rangle\langle\uparrow,0|$. These matrices can be written as
\begin{eqnarray}
\sigma^s \otimes |0\rangle\langle 0| , \quad s=x,y,z,
\label{slow.spin}
\end{eqnarray}
where $\sigma^s$ are the Pauli matrices and $|0\rangle\langle
0|$ is the projection operator onto the $q=0$ vibron state.
The expression (\ref{slow.spin}) shows that the
slowest deviations represent spin polarizations in the $x$, $y$, and $z$
direction in the singly occupied sector. Evidently, spin polarizations decay
most slowly. We will return to this point below.

\begin{figure}[tbh]
\centerline{\includegraphics[width=3.60in,clip]{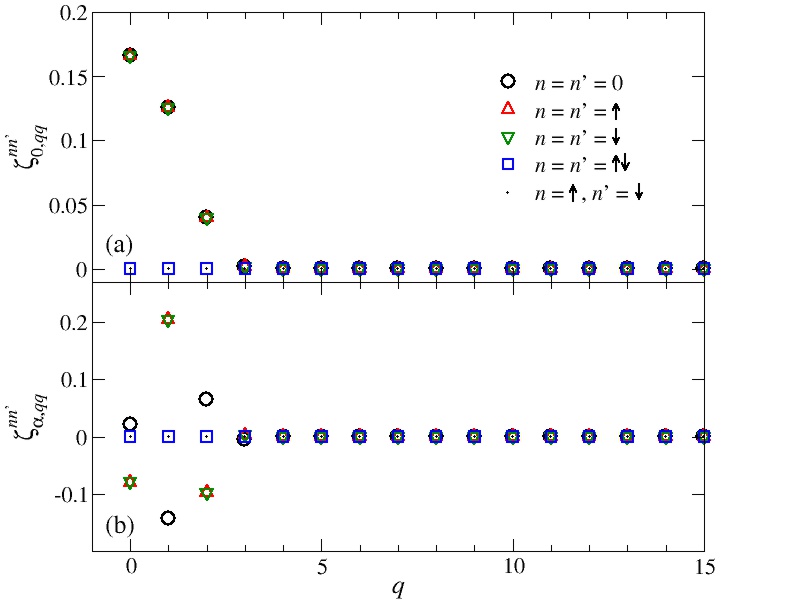}}
\caption{Secular components $\zeta^{nn'}_{\alpha,qq}$ of (a) the
stationary density matrix and (b) the slowest deviation for bias voltage
$eV=3$ in units of the vibron energy $\hbar\omega_v=1$. The other
parameters are the same as in Figs.\ \ref{uli1} and \ref{fig1}.}
\label{states1}
\end{figure}

\begin{figure}[tbh]
\centerline{\includegraphics[width=3.60in,clip]{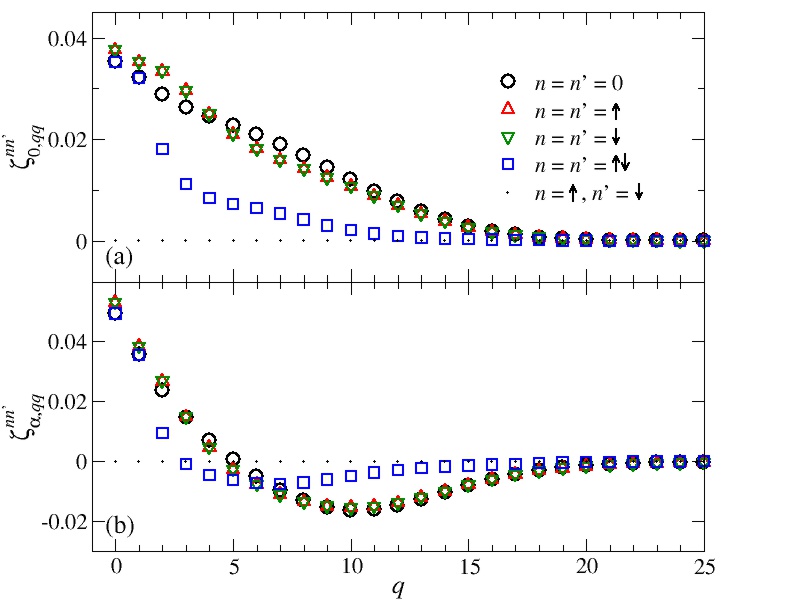}}
\caption{Secular components $\zeta^{nn'}_{\alpha,qq}$ of (a) the
stationary density matrix and (b) the slowest deviation for bias voltage
$eV=11$ in units of the vibron energy $\hbar\omega_v=1$. The other
parameters are the same as in Figs.\ \ref{uli1} and \ref{fig1}.}
\label{states2}
\end{figure}

At the first step in Fig.\ \ref{uli1}, where $eV/2 \approx \hbar\omega_v$,
there is a crossing in the spectrum in Fig.\ \ref{fig1}(a) and we thus expect
the deviation with the smallest decay rate to change in character. Figure
\ref{states1} shows the
secular components of the stationary density matrix and of the slowest
deviation at $eV=3\,\hbar\omega_v$. Compared to $V=0$, the stationary state
obtains finite probabilities for low-lying vibron excitions in the
sectors of electronic occupation numbers $0$ and $1$. The deviation with the
smallest decay rate is now non-degenerate and the large components
$\zeta^{nn'}_{\alpha,qq}$ change sign when the occupation changes between $0$
and $1$ and also when $q$ is increased by unity. How can we understand this?
At $eV=3\,\hbar\omega_v$, $q$ can increase at most by unity in a
sequential-tunneling event. Sequential tunneling also changes the
occupation by $\pm 1$. The slowest deviation is dominated by an imbalance
between the probabilities of the microstates $|0,0\rangle$,
$|{\uparrow},1\rangle$,
$|{\downarrow},1\rangle$, $|0,2\rangle$ on the one hand and of
$|{\uparrow},0\rangle$, $|{\downarrow},0\rangle$, $|0,1\rangle$,
$|{\uparrow},2\rangle$,
$|{\downarrow},2\rangle$ on the other. This imbalance relaxes slowly because
endothermal transitions between any microstate from one class and any
microstate from the other are thermally suppressed.

At the third step at $eV\approx 6\,\hbar\omega_v$, there is another crossing
in Fig.\ \ref{fig1}(a).
Figure \ref{states2} shows the secular components of the stationary
density matrix and of the slowest deviation at $eV=11\,\hbar\omega_v$.
The stationary state now contains highly excited vibrons. Also, the
probabilities of doubly occupied states are comparable to those of empty and
singly occupied states. The slowest deviation is non-degenerate and mainly
involves a transfer of weight between weakly excited vibron
states with $q\lesssim 5$ and highly excited states with $q\gtrsim 5$.
The significance of the number of $5$ becomes clear by inspecting the spectra
in Fig.\ \ref{fig1}: At $eV=11\,\hbar\omega_v$, $q$ can increase by at most
$5$ in a sequential-tunneling event, whereas any decrease is possible.
The deviation sketched in Fig.\ \ref{states2}(b) is mainly an imbalance
between vibron states that differ in $q$ by more than $5$. Such a deviation is
slow to relax by endothermal sequential tunneling since the relaxation requires
more than one transition. We have checked this interpretation by following the
slowest deviation to higher voltages. It retains
its character but the zero of $\zeta^{nn}_{\alpha,qq}$ shifts to higher $q$ (not
shown). This is expected since for higher voltages larger changes of $q$ in a
single transition become possible. Figure \ref{fig1}(a) shows that the slowest
modes become slower with increasing voltage, due to the decrease of
Franck-Condon matrix elements $F_{qq'}$ for larger $|q-q'|$.

\begin{figure}[tbh]
\centerline{\includegraphics[width=3.60in,clip]{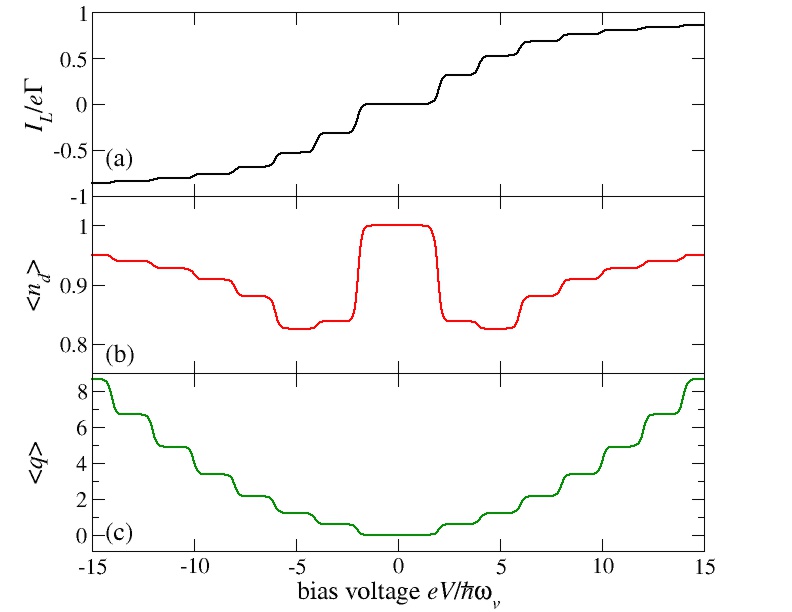}}
\caption{(Color online) (a) Current, (b) electronic occupation, and (c)
vibron excitation as functions of bias voltage $eV$ for
relatively small electron-vibron coupling $\lambda=1$, on-site energy
$\epsilon_d=0$, Hubbard interaction $U=6$, and thermal energy
$k_BT=0.05$. All energies are given in units of the vibron energy
$\hbar\omega_v=1$.}
\label{uli}
\end{figure}

\begin{figure}[tbh]
\centerline{\includegraphics[width=3.40in,clip]{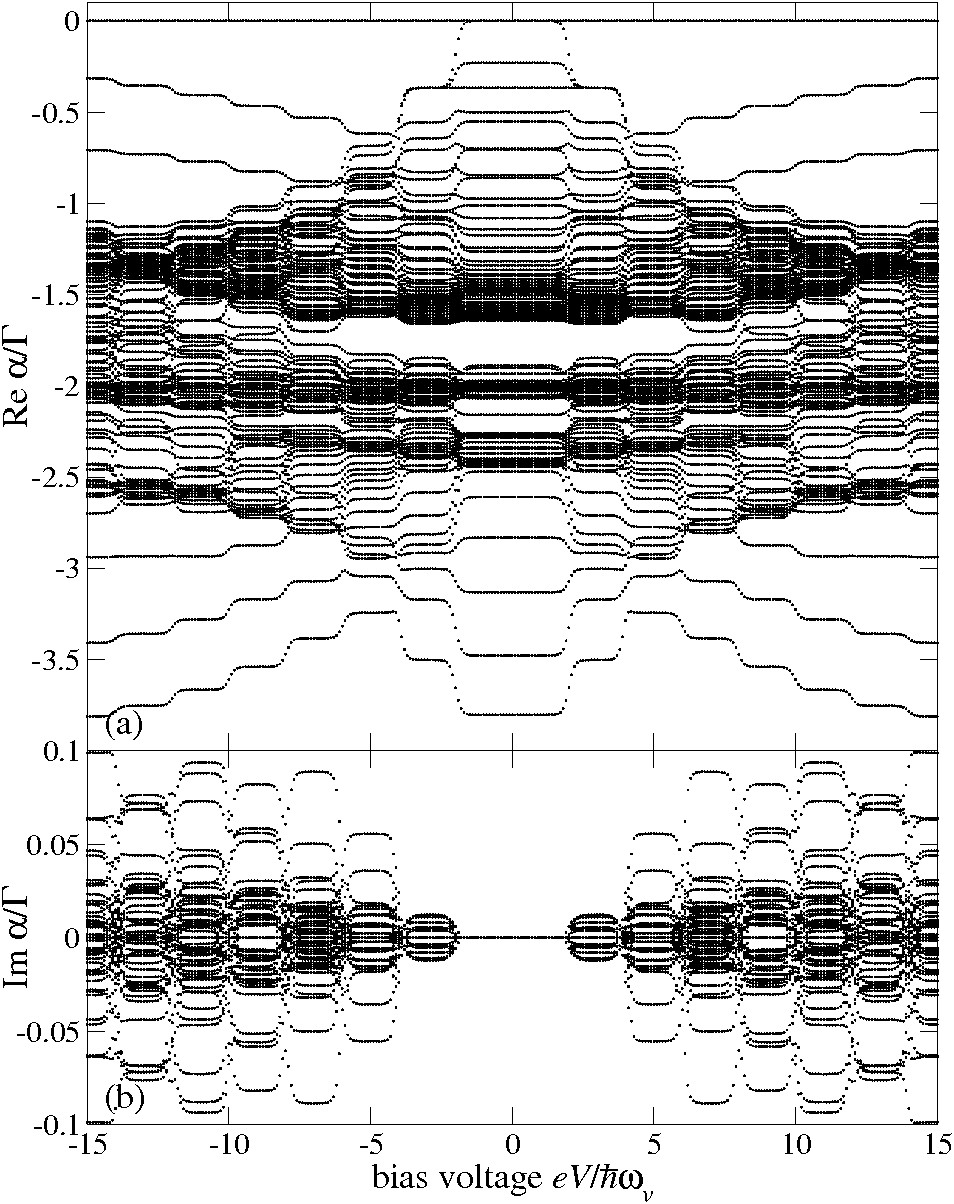}}
\caption{(a) Real and (b) imaginary parts of the eigenvalues of the
transition-rate matrix as functions of the applied bias voltage $eV$. The
parameters are the same as in Fig.\ \ref{uli}.}
\label{fig3}
\end{figure}

\subsection{Coulomb blockade}

If the molecular energy level $\epsilon_d$ is detuned from resonance, there is
a non-zero excitation energy between states with different occupation numbers
and the current is suppressed at small bias voltages. First, we consider the
case that the molecular energy level $\epsilon_d$ lies below the Fermi energy of
the leads at zero bias but the addition energy
$\epsilon_d+U$ for a second electron lies above. In this Coulomb-blockade
regime, the current is suppressed by the Coulomb repulsion $U$. For this
regime, Fig.\
\ref{uli} shows the current $I_L$, the electronic occupation $\langle
n_d\rangle$, and the vibron excitation $\langle q\rangle$
as functions of bias voltage in the stationary state for relatively small
electron-vibron coupling $\lambda=1$. Compared to the transmitting regime, we
observe Coulomb blockade for small bias voltages ($|eV| \lesssim
2\hbar\omega_v$),
where all three observables are
approximately constant and the stationary state is an equal mixture of the
degenerate singly occupied ground states $|{\uparrow},0\rangle$ and
$|{\downarrow},0\rangle$ with exponentially small corrections. When the
bias voltage reaches a certain threshold, electrons can tunnel out of the
molecule so that the average occupation number
decreases, see Fig.\ \ref{uli}(b), and the current sets in, see
Fig.\ \ref{uli}(a). For higher voltages, also doubly occupied states occur with
significant probability and the average occupation number increases again.
Similarly to the transmitting regime, the current increases in steps due to
inelastic tunneling under excitation of vibrons, as seen from the
increase in $\langle q\rangle$ in Fig.\ \ref{uli}(c).

Figure \ref{fig3} shows the real and imaginary parts of the eigenvalues for the
same parameters used in Fig.\ \ref{uli}. Clearly, as the system enters the
Coulomb blockade, a non-zero real eigenvalue becomes very small.
This is a threefold degenerate eigenvalue corresponding to
deviations of the form (\ref{slow.spin}). Thus in the Coulomb blockade,
the spin polarization in the singly occupied sector decays very slowly. This is
easy to understand: An electron has
to tunnel out of the molecule and another electron with opposite spin has to
tunnel in (or vice versa) to relax the spin. But the first tunneling process is
thermally suppressed by the exponentially small tail of the Fermi
function. If we were to include
higher orders in $t_{L,R}$ in the calculation, eigenvalues with exponentially
small leading-order contribution would generically obtain a contribution of
order $|t_{L,R}|^4$ that is not exponentially suppressed but is still small as
long as the pertubative expansion in $t_{L,R}$ is justified.
We have shown above that the same spin deviations still decay
slowly, although not with exponentially suppressed rate, in the transmitting
regime. Note that the inclusion of the off-diagonal secular components of
$\rho_\mathrm{mol}$ is necessary to obtain the correct spin symmetry and
degeneracy of these slow modes.

\begin{figure}[tbh]
\centerline{\includegraphics[width=3.60in,clip]{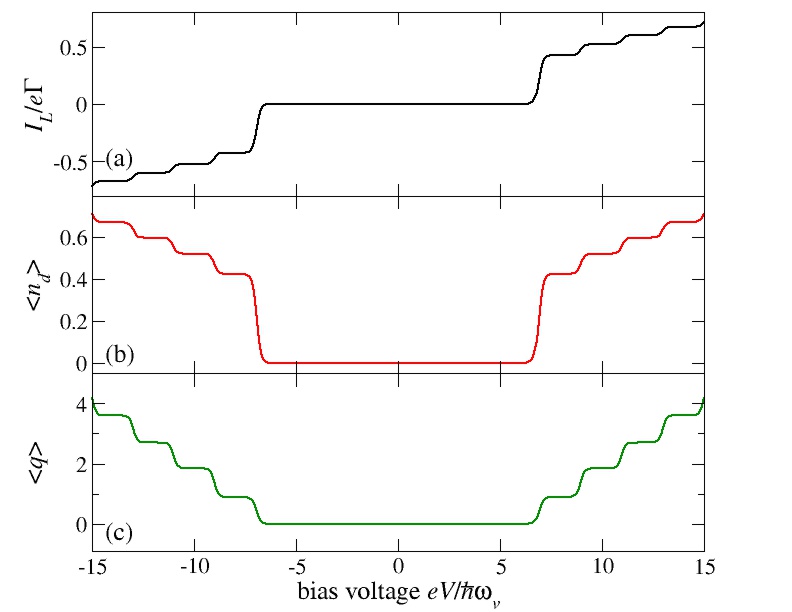}}
\caption{(Color online) (a) Current, (b) electronic occupation, and (c) vibron
excitation as functions of bias voltage $eV$ for
relatively small electron-vibron coupling $\lambda=1$, on-site energy
$\epsilon_d=4.5$, Hubbard interaction $U=6$, and thermal energy
$k_BT=0.05$. All energies are given in units of the vibron energy
$\hbar\omega_v=1$.}
\label{uli_a}
\end{figure}

\begin{figure}[tbh]
\centerline{\includegraphics[width=3.40in,clip]{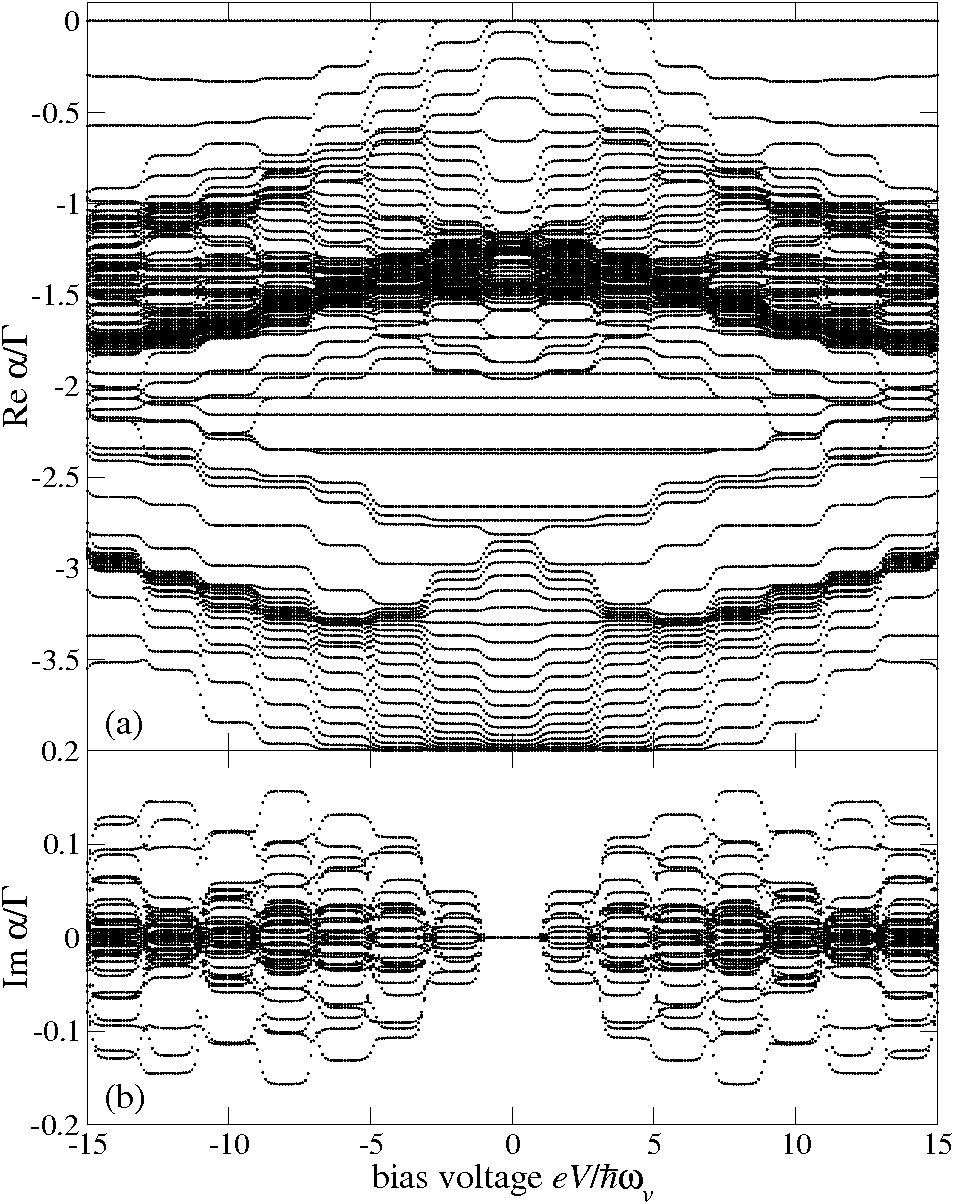}}
\caption{(a) Real and (b) imaginary parts of the eigenvalues of the
transition-rate matrix as functions of the applied bias voltage $eV$. The
parameters are the same as in Fig.\ \ref{uli_a}.}
\label{fig3_a}
\end{figure}

Next, we turn to the two regimes where both $\epsilon_d$ and
$\epsilon_d+U$ lie either above or below the Fermi energy of the leads. Then,
the molecular orbital in the stationary state is either predominantly empty or
doubly occupied, respectively. The two regimes are related to each other by a
particle-hole transformation so that the transport properties are very similar.
In these regimes it is the single-particle energy rather than the Coulomb
interaction that suppresses sequential tunneling. We nevertheless continue to
use the term ``Coulomb blockade''.
For a predominantly empty or doubly occupied molecular orbital, the molecular
spin is essentially zero and its relaxation should not be important for the
dynamics, like it was in the previous case. We plot the stationary current,
electronic occupation, and vibron excitation as functions of the
bias voltage for $\epsilon_d=4.5\, \hbar\omega_v$ in Fig.\ \ref{uli_a}.
There is now a broad regime at low bias voltage where the
molecular orbital is essentially empty. Singly occupied states
become available above the Coulomb-blockade threshold so that a
sequential-tunneling current sets in. Vibrons start to be excited at the same
point since an electron tunneling out of the molecule has sufficient excess
energy to excite the vibration. The corresponding eigenvalue spectra are
plotted in Fig.\ \ref{fig3_a}. It is striking that in this case
no eigenvalue becomes small right at the threshold at $eV\approx
\pm 7\,\hbar\omega_v$---the gap
in the spectrum persists into the Coulomb-blockade regime. An eigenvalue
approaches zero, i.e., the gap closes, only at a voltage of $eV\approx \pm
5\,\hbar\omega_v$. For smaller voltages, even deeper in the Coulomb-blockade
regime, there are
additional transitions where further eigenvalues become small. Note that the
stationary observables in Fig.\ \ref{uli_a} are all exponentially
suppressed here.

\begin{figure}[tbh]
\centerline{\includegraphics[width=3.60in,clip]{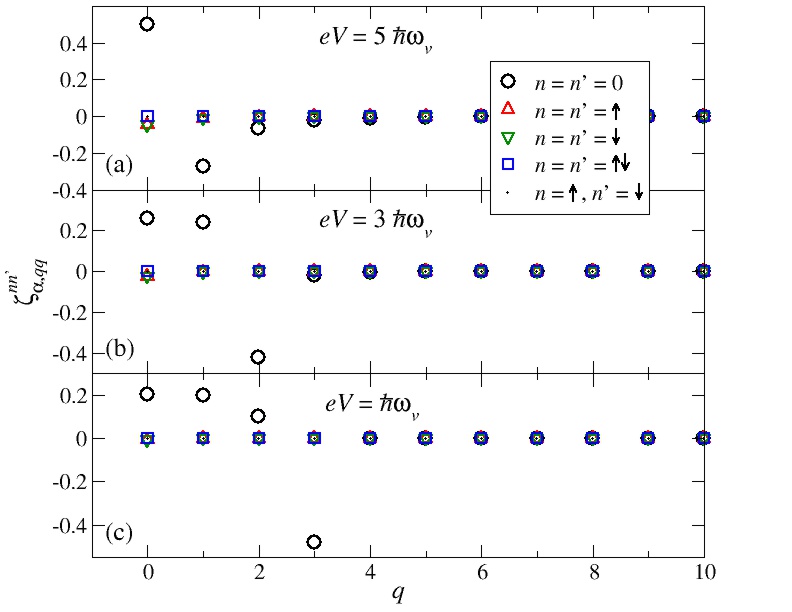}}
\caption{Secular components $\zeta^{nn'}_{\alpha,qq}$ of deviations becoming
slow within the Coulomb-blockade regime in Figs.\ \ref{uli_a} and \ref{fig3_a}.
Panels (a), (b), (c) show the modes becoming slow at $eV=5\,\hbar\omega_v$,
$3\,\hbar\omega_v$, $\hbar\omega_v$, respectively.
The other parameters are the same as in Figs.\ \ref{uli_a} and \ref{fig3_a}.}
\label{states3}
\end{figure}

The stationary state is of course dominated by $|0,0\rangle$ throughout the
blockade regime. We now analyze the deviations that become slow as the voltage
is lowered. At $eV\approx \pm 5\,\hbar\omega_v$, a non-degenerate mode becomes
slow that mainly involves transfer of weight between $|0,0\rangle$ and
excited vibrational, and to a lesser extend electronic, states.
This slow mode is sketched in Fig.\ \ref{states3}(a).
Below the voltage $eV\approx \pm 5\,\hbar\omega_v$, the
excited-state-to-excited-state transitions from $|0,q+1\rangle$ to
$|{\uparrow},q\rangle$ and $|{\downarrow,}q\rangle$ become suppressed. In
particular, the only rapid decay channel of the state $|0,1\rangle$ (to
$|{\uparrow},0\rangle$ and $|{\downarrow},0\rangle$ and then to
$|0,0\rangle$) is suppressed. Therefore, the slowest deviation mostly involves
transfer of weight between $|0,1\rangle$ and $|0,0\rangle$.

Next, at $eV\approx \pm 3\,\hbar\omega_v$, another non-degenerate mode becomes
slow. It is sketched in Fig.\ \ref{states3}(b). This mode involves a transfer
of weight between the \emph{two} lowest vibrational states $|0,0\rangle$,
$|0,1\rangle$ on the one hand and mainly the next state $|0,2\rangle$ on the
other. It becomes slow because below this voltage also the decay of
$|0,2\rangle$ is suppressed. Analogously, at $eV\approx \hbar\omega_v$, a
further mode sketched in Fig.\ \ref{states3}(c) becomes slow due to the
suppression of the decay of $|0,3\rangle$. If we would increase the on-site
energy further by means of a gate voltage, we expect more and more slow modes to
appear.

\begin{figure}[tbh]
\centerline{\includegraphics[width=3.60in,clip]{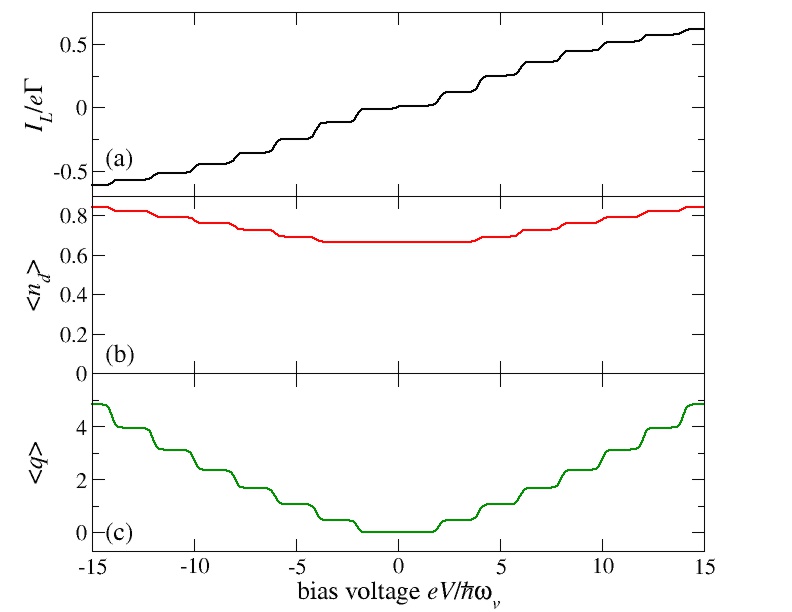}}
\caption{(Color online) (a) Current, (b) electronic occupation, and (c) vibron
excitation as functions of bias
voltage $eV$ for intermediate electron-vibron coupling $\lambda=2$, on-site
energy $\epsilon_d=4$, Hubbard interaction $U=12$, and thermal energy
$k_BT=0.05$. All energies are given in units of the vibron energy
$\hbar\omega_v=1$.}
\label{fig8}
\end{figure}

\begin{figure}[tbh]
\centerline{\includegraphics[width=3.40in,clip]{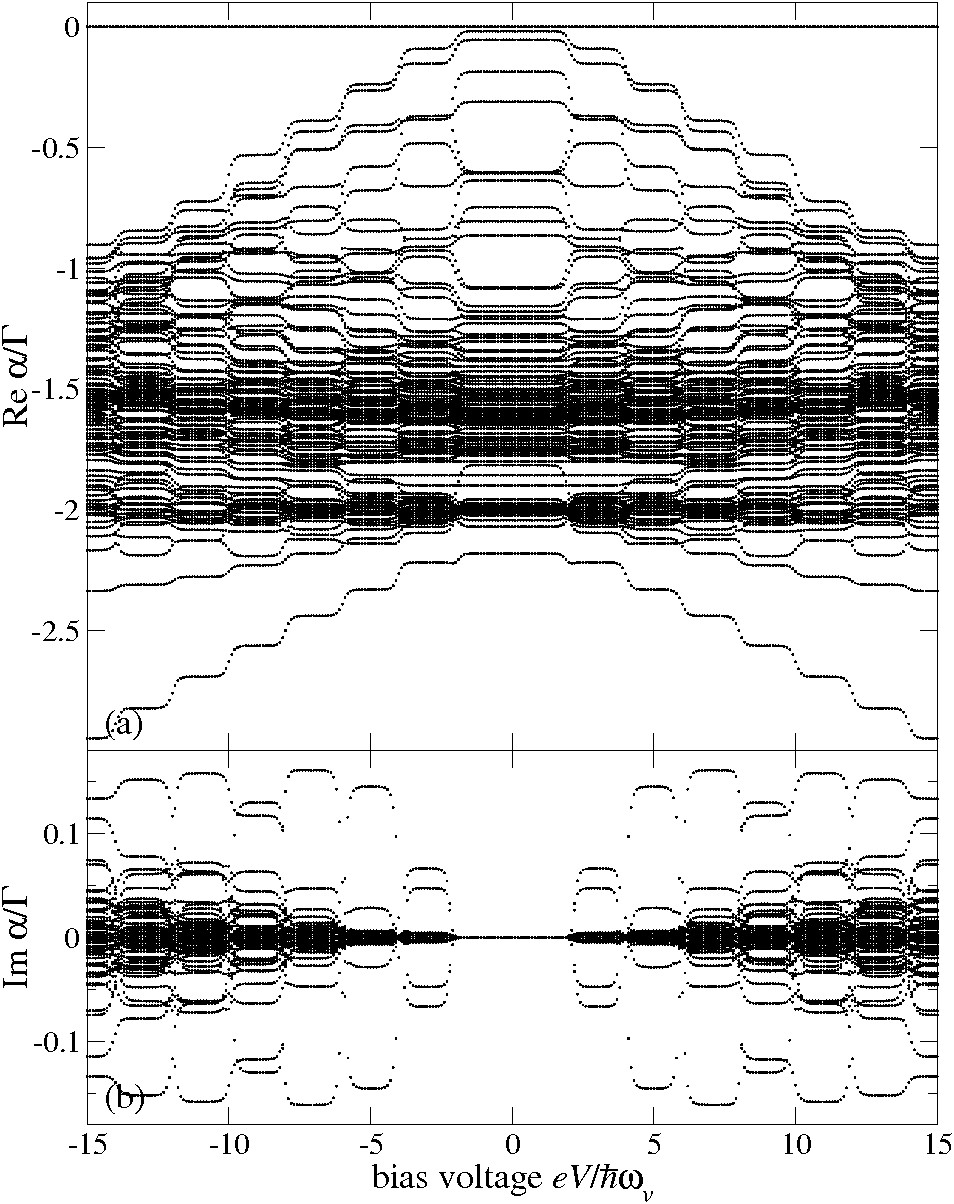}}
\caption{(a) Real and (b) imaginary parts of the eigenvalues of the
transition-rate matrix as functions of the applied bias voltage $eV$. The
parameters are the same as in Fig.\ \ref{fig8}. In particular, the
electron-vibron coupling is $\lambda=2$.}
\label{fig5}
\end{figure}

\subsection{Franck-Condon blockade}

In this subsection, we turn to the signatures of Franck-Condon blockade in the
spectra. Like for the transmitting regime, we tune the effective on-site energy
$\epsilon_d-\lambda^2\hbar\omega_v$ to
zero. Then resonant tunneling is possible at $V=0$ and any suppression is due
to Franck-Condon blockade. Figure \ref{fig8} shows the
stationary current, electronic occupation, and vibron excitation as
functions of the
bias voltage for intermediate electron-vibron coupling
$\lambda=2$. We choose a larger Hubbard interaction $U=12\,\hbar\omega_v$ since
for the previously used value of $U=6\,\hbar\omega_v$, the effective
interaction in Eq.\ (\ref{eigen}) would become attractive. The main effect of
the stronger electron-vibron coupling is the suppression of the zero-bias
current step in Fig.\ \ref{fig8}(a).

The corresponding eigenvalue spectra are plotted in Fig.\ \ref{fig5}. We find
a smaller gap at low bias voltage, compared to the case of $\lambda=1$ shown in
Fig.\ \ref{fig1}. At $V=0$, the smallest eigenvalue is threefold degenerate
and corresponds to spin imbalances of the form (\ref{slow.spin}). The slowest
deviations are thus the same as for the transmitting regime,
but their decay rate has become even smaller. At first glance, it
might be surprising that the
enhancement of electron-vibron coupling leads to suppressed \emph{spin}
relaxation. The reason is that in order for the spin to relax,
electrons have to tunnel in and out of the molecule. At low
voltage, the only available transitions are between $|0,0\rangle$ on the one
hand and $|{\uparrow},0\rangle$ and $|{\downarrow},0\rangle$ on the other. But
these transitions are now suppressed by the small Franck-Condon matrix element
$F_{00} = e^{-\lambda^2/2}$.
The next eigenvalue, which is comparable in magnitude, is not degenerate
and corresponds to an imbalance between the empty and singly occupied states.
It becomes slow for the same reason.

\begin{figure}[tbh]
\centerline{\includegraphics[width=3.60in,clip]{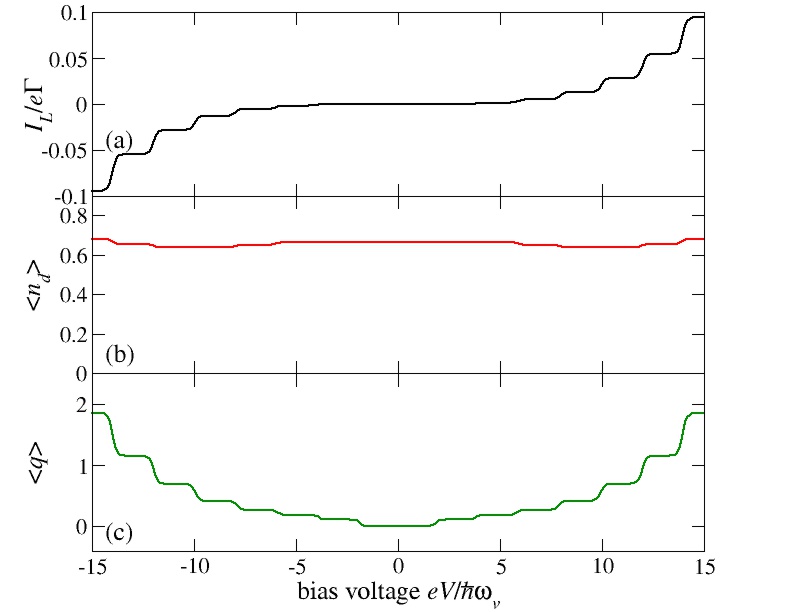}}
\caption{(Color online) (a) Current, (b) electronic occupation, and (c)
vibron excitation as functions of bias
voltage $eV$ for strong electron-vibron coupling $\lambda=4$, on-site
energy $\epsilon_d=16$, Hubbard interaction $U=40$, and thermal energy
$k_BT=0.05$. All energies are given in units of the vibron energy
$\hbar\omega_v=1$.}
\label{fig8_a}
\end{figure}

\begin{figure}[tbh]
\centerline{\includegraphics[width=3.40in,clip]{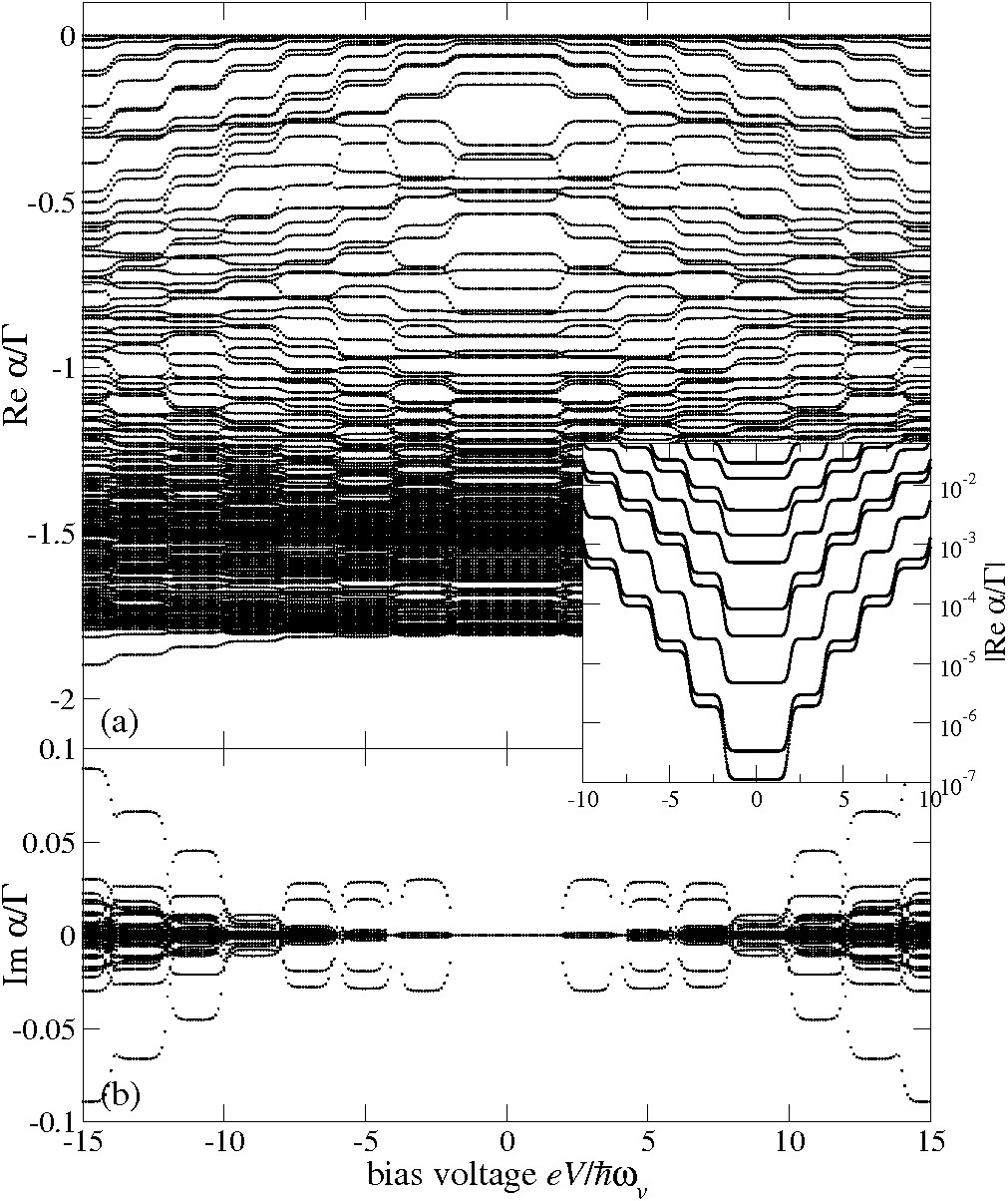}}
\caption{(a) Real and (b) imaginary parts of the eigenvalues of the
transition-rate matrix as functions of the applied bias voltage $eV$. Inset:
Absolute value of the smallest non-vanishing eigenvalues on a logarithmic scale.
The parameters are the same as in Fig.\ \ref{fig8_a}. In particular, the
electron-vibron coupling is $\lambda=4$.}
\label{fig5_a}
\end{figure}

Finally, we turn to the case of even stronger electron-vibron coupling
$\lambda$. The effective on-site energy $\epsilon_d-\lambda^2\hbar\omega_v$
is again tuned to zero. Figure \ref{fig8_a} shows the
stationary current, electronic occupation, and vibron excitation as
functions of the
bias voltage for strong electron-vibron coupling $\lambda=4$ and
$U=40\,\hbar\omega_v$. Due to the large value of $U$, a larger cutoff
$q_\mathrm{max}=50$ is chosen here. Note the current scale in Fig.\
\ref{fig8_a}(a): The current is strongly reduced in magnitude for all voltages,
in particular for small ones, by the Franck-Condon
blockade.\cite{KoO05,KRO05,KOA06,DGR06,Wegewijs,HuB09} In this regime, the
voltage dependence of the occupation number and of the vibron excitation are
also suppressed.
The corresponding eigenvalue spectra are plotted in Fig.\ \ref{fig5_a}.
The typical real and imaginary parts have become smaller
and the gap is completely filled in at all bias voltages shown here. Thus there
are slow modes in the whole voltage range. At least at small voltages, the
character of the slowest modes is the same as for $\lambda=2$, though: The most
long-lived deviations are spin and charge imbalances, the decay of which is
suppressed by small Franck-Condon matrix elements.

The inset in Fig.\ \ref{fig8_a} shows details on the small real parts on a
logarithmic scale. The small real parts roughly follow a log-uniform
distribution for a
certain range of rates. Within this range, the probability density function
is approximately $P(|\mathrm{Re}\,\alpha|) \sim 1/|\mathrm{Re}\,\alpha|$, i.e.,
it is scale-invariant.
The uniform distribution of $\ln |\mathrm{Re}\,\alpha|$ is caused by
the approximately exponential dependence of the Franck-Condon matrix elements
$F_{qq'}$ on $q$ and $q'$ for $q,q'\ll \lambda^2$.
The approximate scale-invariance of the distribution of small rates implies
that the dynamics of the system within a certain time window is also
scale-invariant. This is consistent with the approximate self-similarity of the
time-dependent transport found by Monte Carlo simulations.\cite{KoO05,KRO05}

\section{Summary and Conclusions}
\label{sec:conc}

In the present paper, we have studied the eigenvalue spectrum of the
transition-rate matrix in the ME for a molecular quantum dot coupled to metallic
leads. The relaxational and oscillatory dynamics of deviations of the system
from the stationary state are characterized by the real and imaginary parts of
these eigenvalues, respectively. We have mainly considered the small
eigenvalues, which describe the slow dynamics. Conceptually, this is similar to
analyzing the spectrum of low-lying eigenenergies of a \emph{Hamiltonian}
system. We have applied this idea to a molecular transistor with an electronic
orbital coupled to a vibrational mode.

The spectra differ qualitatively between a transmitting device, a molecule in
the Coulomb-blockade regime, and a molecule in the Franck-Condon-blockade
regime. We demonstrate that the character of deviations from the stationary
state can be analyzed by considering the large components of the corresponding
eigenvectors. Some of the deviations with the smallest decay
rates represent non-zero spin polarizations of the molecule. They occur in
groups of three degenerate modes corresponding to polarizations in the
\textit{x}, \textit{y}, and \textit{z} direction. In order to obtain these
modes, all secular components of the reduced density matrix have to be included.

In the transmitting regime, the spectrum has a gap for any bias voltage,
i.e., there are no slowly decaying deviations on the scale of the
sequential-tunneling rate $\Gamma$. In the Coulomb-blockade regime with
predominantly single occupation of the molecular orbital, the gap in the
spectrum closes since relaxation of the electronic spin becomes slow. If instead
the molecular orbital is predominantly empty or doubly occupied, there is no
finite spin polarization and thus these slow modes do not exist. In these cases,
the gap persists into the Coulomb-blockade regime.
However, deep within these regimes the gap closes and more and
more modes become
slow at consecutive steps. These modes become slow since
excited-state-to-excited-state transitions are thermally suppressed. The
dynamics here contains additional information not accessible by
observables in the stationary state, which show an exponentially
suppressed voltage dependence.

For stronger electron-vibron couping we find that the gap becomes small even if
resonant tunneling is energetically possible, since certain transition rates are
suppressed by small Franck-Condon matrix elements. In the strong
Franck-Condon-blockade regime, the gap closes over a broad range of bias
voltages since many deviations now decay slowly. We
also find an approximately scale-invariant distribution of the slowest rates,
consistent with the previously observed self-similar dynamics in real
time.\cite{KoO05,KRO05}

In the present paper, we have concentrated on the stationary and long-lived
states. The spectra obtained in this paper show additional structure that we
have not discussed, suggesting that much more information can be extracted from
the spectra and the eigenmodes.

\acknowledgments

We would like to thank S. Lange and T. Ludwig for helpful discussions.
Financial support by the Deutsche Forschungsgemeinschaft is gratefully
acknowledged.


\begin{thebibliography}{99}


\bibitem{Rat02}M. A. Ratner, Materials Today \textbf{5} (2), 20 (2002); K. S.
Kwok and J. C. Ellenbogen, \textit{ibid.}, p.\ 28.

\bibitem{XuR06}Y. Xue and M. A. Ratner, in \textit{Nanotechnology: Science and
Computation}, edited by J. Chen, N. Jonoska, and G. Rozenberg (Springer, Berlin,
2006), p.\ 215.

\bibitem{ABV06}C. H. Ahn, A. Bhattacharya, M. Di Ventra, J. N. Eckstein, C. D.
Frisbie, M. E. Gershenson, A. M. Goldman, I. H. Inoue, J. Mannhart, A. J.
Millis, A. F. Morpurgo, D. Natelson, and J.-M. Triscone, Rev.\ Mod.\ Phys.\
\textbf{78}, 1185 (2006).

\bibitem{GRN07}M. Galperin, M. A. Ratner, and A. Nitzan, J. Phys.: Condens.\
Matter \textbf{19}, 103201 (2007).

\bibitem{Andergassen}S.\ Andergassen, V.\ Meden, H.\ Schoeller, J.\
Splettstoesser, and M. R. Wegewijs, Nanotechnology {\bf 21}, 272001 (2010).

\bibitem{ZiP11}N. A. Zimbovskaya and M. R. Pederson, Phys.\ Rep.\
\textbf{509}, 1 (2011).

\bibitem{Park}H.\ Park, J.\ Park, A. K. L. Lim, E. H.\ Anderson, A. P.\
Alivisatos, and P. L.\ McEuen, Nature {\bf 407}, 57 (2000).

\bibitem{OOS07}E. A. Osorio, K. O'Neill, N. Stuhr-Hansen, O. F. Nielsen, T.
Bj\o{}rnholm, and H. S. J. van der Zant, Adv.\ Mater.\ \textbf{19}, 281 (2007).

\bibitem{Smit}R. H. M.\ Smit, Y.\ Noat, C.\ Untiedt, N. D.\ Lang, M. C.\ van
Hemert, and J. M.\ van Ruitenbeek, Nature {\bf 419}, 906 (2002).

\bibitem{Yu}L. H.\ Yu, Z. K.\ Keane, J. W.\ Ciszek, L.\ Cheng, M. P.\
Stewart, J. M.\ Tour, and D. Natelson, Phys.\ Rev.\ Lett.\ {\bf 93}, 266802
(2004).

\bibitem{RLM10}J. Repp, P. Liljeroth, and G. Meyer, Nature Phys.\
\textbf{6}, 975 (2010).

\bibitem{Fleb}K.\  Flensberg, Phys.\ Rev.\ B {\bf 68}, 205323 (2003).

\bibitem{Braig}S.\ Braig and K.\ Flensberg, Phys.\ Rev.\ B {\bf 68}, 205324
(2003).

\bibitem{KoO05}J. Koch and F. von Oppen, Phys.\ Rev.\ Lett.\ \textbf{94},
206804 (2005).

\bibitem{KRO05}J. Koch, M. E. Raikh, and F. von Oppen, Phys.\ Rev.\ Lett.\
\textbf{95}, 056801 (2005).

\bibitem{KOA06}J. Koch, F. von Oppen, and A. V. Andreev, Phys.\ Rev.\ B
\textbf{74}, 205438 (2006).

\bibitem{DGR06}A. Donarini, M. Grifoni, and K. Richter, Phys.\ Rev.\ Lett.\
\textbf{97}, 166801 (2006).

\bibitem{Wegewijs}M.\ Leijnse and M. R.\ Wegewijs, Phys.\ Rev.\ B {\bf
78}, 235424 (2008).

\bibitem{HuB09}H. H\"ubener and T. Brandes, Phys.\ Rev.\ B \textbf{80}, 155437
(2009).

\bibitem{Leturcq}R.\ Leturcq, C.\ Stampfer, K.\ Inderbitzin, L.\ Durrer, C.\
Hierold, E.\ Mariani, M. G.\ Schultz, F.\ von Oppen, and K. Ensslin, Nature
Phys.\ {\bf 5}, 327 (2009).

\bibitem{CML10}F. Cavaliere, E. Mariani, R. Leturcq, C. Stampfer, and M.
Sassetti, Phys.\ Rev.\ B \textbf{81}, 201303(R) (2010).

\bibitem{Mitra}A.\ Mitra, I.\ Aleiner, and A.\ J.\ Millis, Phys.\ Rev.\ B {\bf
69}, 245302 (2004).

\bibitem{Elste}F.\ Elste, G.\ Weick, C. \ Timm, and F. von Oppen, Appl.\ Phys.\
A \textbf{93}, 345 (2008).

\bibitem{HBT09}R.\ H\"artle, C.\ Benesch, and M.\ Thoss, Phys.\ Rev.\ Lett.\
{\bf 102}, 146801 (2009).

\bibitem{ScS94}H. Schoeller and G. Sch\"on, Phys.\ Rev.\ B \textbf{50}, 18436
(1994).

\bibitem{KSS95}J. K\"onig, H. Schoeller, and G. Sch\"on, Europhys.\ Lett.\
\textbf{31}, 31 (1995);
J. K\"onig, H. Schoeller, and G. Sch\"on, Phys.\ Rev.\ Lett.\
\textbf{76}, 1715 (1996);
J. K\"onig, J. Schmid, H. Schoeller, and G. Sch\"on, Phys.\ Rev.\ B \textbf{54},
16820 (1996).

\bibitem{Timm}C.\ Timm, Phys.\ Rev.\ B.\ {\bf 77}, 195416 (2008).

\bibitem{Blum}K. Blum, \textit{Density Matrix Theory and Applications} (Plenum,
New York, 1981).

\bibitem{Koller}S.\ Koller, M.\ Grifoni, M.\ Leijnse, and M. R.\ Wegewijs,
Phys.\ Rev.\ B \textbf{82}, 235307 (2010).

\bibitem{SKB12}G. Schaller, T. Krause, T. Brandes, and M. Esposito,
arXiv:1206.3960.

\bibitem{DYG12}A. Donarini, A. Yar, and M. Grifoni, arXiv:1205.4927v1.


\bibitem{GlS88}L. I. Glazman and R. I. Shekhter, Sov.\ Phys.\ JETP \textbf{67},
163 (1988).

\bibitem{WJW88}N. S. Wingreen, K. W. Jacobsen, and J. W. Wilkins, Phys.\ Rev.\
Lett.\ \textbf{61}, 1396 (1988).

\bibitem{And61}P. W. Anderson, Phys.\ Rev.\ \textbf{124}, 41 (1961).

\bibitem{BrP02}H.-P. Breuer and F. Petruccione, \textit{The Theory of Open
Quantum Systems} (Oxford University Press, Oxford, 2002).

\bibitem{Nak58}S. Nakajima, Prog.\ Theor.\ Phys.\ \textbf{20}, 948 (1958).

\bibitem{Zwa60}R. Zwanzig, J. Chem.\ Phys.\ \textbf{33}, 1338 (1960); Physica
(Amsterdam) \textbf{30}, 1109 (1964).

\bibitem{ToM76}M. Tokuyama and H. Mori, Prog.\ Theor.\ Phys.\ \textbf{55}, 411
(1976).

\bibitem{STH77}N. Hashitsume, F. Shibata, and M. Shing\={u}, J. Stat.\ Phys.\
\textbf{17}, 155 (1977); F. Shibata, Y. Takahashi, and N. Hashitsume,
\textit{ibid.}\ \textbf{17}, 171 (1977).

\bibitem{Tim11}C. Timm, Phys.\ Rev.\ B \textbf{83}, 115416 (2011).

\bibitem{Bre99}P. Br\'emaud, \textit{Markov Chains, Gibbs Fields, Monte Carlo
Simulations and Queues}, Texts in Applied Mathematics
(Springer, New York, 1999), Vol.\ 31.

\bibitem{Muk00}D. Mukamel, in \textit{Soft and Fragile Matter: Nonequilibrium
Dynamics, Metastability and Flow}, edited by M. E. Cates and M. R. Evans
(Institute of Physics Publishing, Bristol, 2000), p.\ 237.

\bibitem{ZiS06}R. K. P. Zia and B. Schmittmann, J. Phys.\ A: Math.\ Gen.\
\textbf{39}, L407 (2006).

\bibitem{timm2}C. Timm, Phys.\ Rev.\ E {\bf 80}, 021140 (2009).

\bibitem{BeP79}A. Berman and R. J. Plemmons, \textit{Non-negative Matrices in
the Mathematical Sciences} (Academic Press, New York, 1979).

\bibitem{ElT05}F. Elste and C. Timm, Phys.\ Rev.\ B \textbf{71}, 155403 (2005).

\bibitem{TiE06}C. Timm and F. Elste, Phys.\ Rev.\ B \textbf{73}, 235304 (2006).

\bibitem{Mathematica}Wolfram Research, Inc., Mathematica, Version 8.0,
Champaign, IL (2010).

\bibitem{StW72}M. M. Sternheim and J. F. Walker, Phys.\ Rev.\ C \textbf{6}, 114
(1972).


\bibitem{BYP05}W. J. Bruno, J. Yang, and J. E. Pearson, PNAS \textbf{102}, 6326
(2005).

\end{thebibliography}
\end{document}